\documentclass[11pt,a4paper]{article}
\RequirePackage{ETHlogo}
\usepackage{graphicx}  
\usepackage{psfrag}
\usepackage{fancyhdr,lastpage}
\usepackage[NewCommands]{ragged2e}
\usepackage{amsmath,amssymb}
\usepackage{latexsym}
\usepackage{url}
\usepackage[colorlinks]{hyperref}
\usepackage[dvips]{color}
\usepackage[sort&compress,numbers]{natbib}
\usepackage{pstricks,pst-node,pst-coil,pst-plot}
\newtheorem{theorem}{Theorem}
\newtheorem{prop}[theorem]{Proposition}

\newtheorem{defn}[theorem]{Definition}

\RequirePackage{fancyhdr}
\long\def\nologohead#1{
    \fancyhead[L]{}
    \fancyhead[C]{  \hrulefill \\[1ex]
        \begin{minipage}[c]{1.0\linewidth}
            \begin{center}
                {\footnotesize{#1}}
            \end{center}
        \end{minipage}\\[-1ex]
        \hrulefill}
    \fancyhead[R]{}
}
\newenvironment{declaration}[1]{\trivlist \item[\hskip \labelsep{\bf #1 }]
\ignorespaces}{\endtrivlist}

\newenvironment{proofof}[1]{\begin{declaration}{#1}}{\end{declaration}}
\newenvironment{proof}{\begin{proofof}{Proof.}}{\end{proofof}}
\nologohead{M.  D. K{\"o}nig, S. Battiston, M. Napoletano, F.
  Schweitzer: \\ On Algebraic Graph Theory and the Dynamics of Innovation
    Networks \\  \emph{Networks and Heterogeneous
    Media} (2007, submitted) \\ See \url{http://www.sg.ethz.ch} for more
  information}
\DeclareMathOperator*{\argmax}{argmax}
\begin{document}

\vspace*{0.2cm}
\begin{center}
  \textbf{\Large On Algebraic Graph Theory and the Dynamics of Innovation
    Networks}\\[5mm]
  {\large M. D. K\"onig$^{\star}$, S. Battiston, M. Napoletano and F.
    Schweitzer}
 
  \emph{Chair of Systems Design, ETH Zurich, Kreuzplatz 5, 8032 Zurich,
    Switzerland}
 
  $^{\star}$ corresponding author, email: \url{mkoenig@ethz.ch} 
\end{center}

\begin{abstract}
  We investigate some of the properties and extensions of a dynamic
  innovation network model recently introduced in
  \citep{koenig07:_effic_stabil_dynam_innov_networ}. In the model, the
  set of efficient graphs ranges, depending on the cost for maintaining a
  link, from the complete graph to the (quasi-) star, varying within a
  well defined class of graphs. However, the interplay between dynamics
  on the nodes and topology of the network leads to equilibrium networks
  which are typically not efficient and are characterized, as observed in
  empirical studies of R\&D networks, by sparseness, presence of clusters
  and heterogeneity of degree. In this paper, we analyze the relation
  between the growth rate of the knowledge stock of the agents from R\&D
  collaborations and the properties of the adjacency matrix associated
  with the network of collaborations. By means of computer simulations we
  further investigate how the equilibrium network is affected by
  increasing the evaluation time $\tau$ over which agents evaluate
  whether to maintain a link or not. We show that only if $\tau$ is long
  enough, efficient networks can be obtained by the selfish link
  formation process of agents, otherwise the equilibrium network is
  inefficient. This work should assist in building a theoretical
  framework of R\&D networks from which policies can be
  derived that aim at fostering efficient innovation networks.\\
  
\end{abstract}

\section{Introduction}
\label{sec:intro}

The field of Network Theory has only recently focused its attention on
the study of dynamic models in which the topology of the network
endogeneously drives the evolution of the network. These models assume
that the evolution of the links in the network is driven by the dynamics
of a state variable, associated to each node, which depends, through the
network, on the state variable of the other nodes
\citep{gross07:_adapt_coevol_networ_review, stewart04:_networ}. Such an
interplay is crucial in many biological systems and especially in
socio-economic systems. In biological systems, a Darwinian selection
mechanism usually works at a global level: for instance in the context of
networks, one can think of a mechanism in which the least fit nodes are
replaced (together with their connections) with new nodes that are
randomly connected to the remaining nodes\citep{jain01,
  jain98:_autoc_sets_growt_compl_evolut,
  caldarelli2007:_extremal_dynamics}. In socio-economic networks, besides
the global selection mechanism, there exists a ``local'' selection
mechanism: the nodes in fact represent agents that form or delete links
with other agents, based on the utility that those links may provide to
them \citep{holme06:_dynam_networ_agent_compet_high,
  carvalho07:_socioec_networ_long_range_inter}.

The foregoing issue has also attracted researchers in computer science
\citep{koutsoupias99:_worst, fabrikant03, corbo05} as well as social
scientists and economists
\citep{jackson96:_strat_model_social_econom_networ,
  goyal00:_Noncooperative_model_network_formation,
  ballester06:_who_networ, haller07:_non_nash,
  haller05:_nash_networ_heter_links}
In particular, the study of networks has become increasingly important in
the literature on R\&D networks \citep{goyal01:_r_d_networ,
  ehrhardt-2006, cowan04:_networ_struc_diffus_knowl}. Here, the evolution
of the network is of interest for the investigation of the efficiency and
stability of networks of agents exchanging knowledge in R\&D
collaborations. In such a context, we have recently introduced a new
model of network evolution in which the topology and the state variable
of nodes co-evolve \citep{koenig07:_effic_stabil_dynam_innov_networ,
  koenig07:_innov_networ_new_approac_model_analy}. The nodes of the
network are associated with a dynamical state variable, representing the
utility of the agent, that depends on the links (the R\&D collaborations)
and the utility of the other agents. The network then evolves according
to a prescribed link formation rule, which, in turn, depends on the
expected increase of utility of the agents.

Independently of the co-evolution of the network and the utility of the
agents, we determine exactly the efficient graphs (in which the aggregate
utility of the agents is maximized) and we show that there are stable
equilibria which are not efficient. This implies that, if the network
evolves through the selfish linking behavior of agents, it may not reach
an efficient equilibrium. This result is of interest for policy design
questions how to establish incentive mechanisms and legal frameworks in
order to help the system to reach its efficient equilibrium.
Interestingly, the model is also able to reproduce some of the main
stylized facts of empirical studies on R\&D networks
\citep{hanaki07:_dynam_r_d_collab_it_indus,
  powell05:_networ_dynam_field_evolut,
  cowan04:_networ_struc_diffus_knowl} - namely that such networks are
sparse, clustered and heterogeneous in degree - and therefore offers a
candidate framework to explain the formation of these networks.

In this paper, we consider the same model of
\citep{koenig07:_effic_stabil_dynam_innov_networ} and we investigate
further some of its properties, in particular, the relation between the
utility of the agents and the properties of the adjacency matrix of the
network.  In addition, here we also introduce a time delay $\tau$ in the
decision about keeping or removing a link and we investigate by means of
computer simulation how the equilibrium reached by the network is
affected by increasing $\tau$. The result is interesting in view of more
realistic models for the design of policies that may facilitate the
formation of efficient innovation networks.
 
The paper is organized as follows. First (sec. \ref{sec:model}), we
introduce the dynamics of knowledge exchange on a static network. Then we
review some results from algebraic graph theory and we discuss their
implications for our model (sec. \ref{sec:matrix_properties}). We proceed
by showing the existence of inefficient equilibria (sec.
\ref{sec:efficiency}-\ref{sec:stability}). We finally report the results
of computer simulations of the evolution of the network, in particular,
with respect to impact of the evaluation time of the links (sec.
\ref{sec:simulation_link_creation}). We finally summarize the results and
draw some conclusions (sec. \ref{sec:discussion}).

\section{Knowledge Dynamics and Utility Function of the Agents}
\label{sec:model}

In this section we describe the dynamics of the state variable of the
nodes in a static network. In section \ref{sec:network_evolution} we
extend our studies to the endogeneous evolution of the network whereby we
introduce the rules for the formation of links.

Consider a set of agents, $N=\{1,...,n\}$, represented as nodes of an
undirected graph $G$, with an associated variable $x_i$ representing the
knowledge of agent $i$. A link $ij$, represents the transfer of knowledge
between agent $i$ and agent $j$.  Knowledge is shared among an
individual's direct and indirect acquaintances and the knowledge level of
an agent is proportional to the knowledge levels of its neighbors.  We
assume that knowledge $\mathbf{x}=(x_1,...,x_n)$ grows, starting from
positive values, $x_i(0)>0$ $\forall i$, according to the following
linear ordinary differential equation
\begin{equation}
  \dot{x}_i = \sum_{j=1}^n a_{ij} x_j \\
  \label{eq:knowledge-growth}
\end{equation}
where $a_{ij}=\{0,1\}$ are the elements of the adjacency matrix
$\mathbf{A}$ of the graph $G$. In vector notation we have
$\dot{\mathbf{x}} = \mathbf{A} \mathbf{x}$. In the following we will use
the terms network and graph as synonyms. 

Similar to \citet{carayol03:_self_organizing_innovation_networks} we
assume that the gross return of agent $i$ is proportional to her
knowledge growth rate, with proportionality constant set to $1$ for sake
of simplicity. We also assume that maintaining a link induces a constant
cost $c \ge 0$ for both agents connected by the link. Therefore the net
return $\rho_i$ of agent $i$ is given by
\begin{equation}
    \rho_i(t) = \frac{\dot{x}_i(t)}{x_i(t)} - c d_i
\label{eq:net-return}
\end{equation}
where $d_i$ denotes the degree of agent $i$.  We assume that the utility
function of an agent in a given network is her asymptotic net return
\mbox{ $ u_i = \lim_{t \to \infty} \rho_i(t)$ }.  As we will show in the
next section, \mbox{ $\lim_{t \to \infty} \frac{\dot{x}_i(t)}{x_i(t)} =
  \lambda_{\text{PF}}(C_i)$ } where $ \lambda_{\text{PF}}$ is the
spectral radius of the block (sub-) matrix in $\mathbf{A}$ corresponding
to the connected component $C_i$ to which $i$ belongs to.  Therefore, the
utility function of agent $i$ in a given network is
\begin{equation}
    u_i(t) =  \lambda_{\text{PF}}(C_i) - c d_i
\label{eq:utility}
\end{equation}
As we will see later on, the evolution of the network stems from each
agent trying independently to increase her utility by forming or deleting
links. Of course when she does so, this affects the utility of the other
agents which will react by forming or removing other links. However,
before describing the evolution of the network, we want to discuss some
implications of our assumptions on the growth of knowledge and the
utility function of the agents.

\section{Knowledge Growth and Properties of the Adjacency Matrix}
\label{sec:matrix_properties}

Since the knowledge dynamics is linear and the utility function is
proportional to the largest real eigenvalue of the graph, there are many
mathematical properties immediately available for the static part of the
model. In this section we review the implications of some well known
results for matrices and graphs on the dynamics of knowledge growth in
the model. We will only focus on undirected graphs and symmetric matrices
respectively.

First of all, since the adjacency matrix in (\ref{eq:knowledge-growth})
is non-negative and in particular it is a Metzler matrix, the vectorial
space $\mathbb{R}^n_+$ is invariant for the linear operator $\mathbf{A}$.
It follows that for non-negative initial values (as assumed in the
model), it is $\mathbf{\dot{x}(t)} \ge 0$ and $\mathbf{x(t)} \ge 0$,
$\forall t>0$ \cite{seneta06:_non_matric_and_markov_chain}. This ensures
the following property:
\begin{prop}
  The values of knowledge $\mathbf{x}$ in (\ref{eq:knowledge-growth}) are
  non-negative for all times.
\label{prop:positive_knowledge}
\end{prop}

For convenience of the reader we report below some facts and definitions
that we need in the succeeding sections.

A walk in the graph is an alternating sequence of nodes and links. The
k-power of the adjacency matrix is related to walks of length $k$ in the
graph. In particular, $\left( \mathbf{A}^k \right)_{ij}$ gives the number
of walks of length $k$ from node $i$ to node $j$
\citep{chris01:_algeb_graph_theor}. A connected component of a graph is a
maximal subgraph in which there exists a walk from every node to every
other node. The graph is connected when the only connected component is
the graph itself. If the adjacency matrix can be decomposed in blocks,
each block corresponds to a connected component.

An $ n\times n$ matrix $A$ is said to be a reducible matrix if and only
if for some permutation matrix $ P$, the matrix $ P^TAP$ is block upper
triangular. If a square matrix is not reducible, it is said to be an
irreducible matrix.  The adjacency matrix of a connected graph is always
irreducible \citep{horn90:_matrix_analy} and in particular it cannot be
decomposed in multiple blocks. Irreducible matrices can be primitive or
cyclic (imprimitive) \citep{seneta06:_non_matric_and_markov_chain}. This
distinction is important because some result about the convergence of the
knowledge values holds only for graphs with primitive adjacency matrix.

For a primitive, non-negative matrix $\mathbf{A}$ it is $\mathbf{A}^k>0$
for some positive integer $k \le (n-1)n^n$ \citep{horn90:_matrix_analy}.
This means that, $\mathbf{A}$ is primitive if, for some $k$, there is a
walk of length $k$ from every node to every other node. Notice that this
definition is a much more restrictive than the one of irreducible (or
connected) graph in which it is required that there exits a walk from
every node to every other node, but not necessarily of the same length. A
graph is said to be primitive if its associated adjacency matrix is
primitive.

The general solution \citep{horn90:_matrix_analy,
  khalil02:_nonlin_system, braun93:_differ_equat_their_applic} of the
system of linear ordinary differential equations in
Eq. (\ref{eq:knowledge-growth}) is
\begin{equation}
  \mathbf{x}(t) = e^{\mathbf{A}t} \mathbf{x}(0)
  \label{eq:matrix_exponential}
\end{equation}
where $\mathbf{x}(0)$ is the initial state and $e^{\mathbf{A}} =
\sum_{n=0}^{\infty} \frac{\mathbf{A}^n}{n!}$ is the matrix exponential.
The matrix exponential can be easily computed in terms of the Jordan form
of the matrix. Once the Jordan form is known $ \mathbf{J} = \mathbf{S}
\mathbf{A} \mathbf{S}^{-1}$, where $S$ is a non-singular matrix, the
matrix exponential is $ e^{\mathbf{A}t} = \mathbf{S} e^{\mathbf{Jt}}
\mathbf{S}^{-1}$. We can then rewrite the solution of the system as
\begin{equation}
  \mathbf{x}(t) = \mathbf{S} e^{\mathbf{Jt}} \mathbf{S}^{-1} \mathbf{x}(0) 
\end{equation}
For each component $i$ of the vector $\mathbf{x}$ we get
\begin{equation}
  x_i(t) = \sum_{j=1}^k p_j(t) e^{\lambda_j t}
  \label{eq:general_solution}
\end{equation}
where $p_j(t)$ is a polynomial in t of degree $\mu_j$ equal to the number
of linearly independent eigenvectors with eigenvalue $\lambda_j$ (its
geometric multiplicity). $\mu_j$ is strictly smaller than the order $m_j$
of the Jordan block corresponding to the eigenvalue $\lambda_j$ of the
matrix $\mathbf{A}$.

If the graph is connected, then the largest real eigenvalue is present in
the exponent for each component $i$ in Eq. (\ref{eq:general_solution}).
Since the matrix $\mathbf{A} \ge 0$ is non-negative, this eigenvalue
coinidides with the spectral radius and with the Perron-Frobenius
eigenvalue, $\lambda_{\text{PF}}$ (see below). Thus, it is
straightforward to see that the ratio $\frac{\dot{x}_i}{x_i}$ is
dominated, for $t \to \infty$, by $\lambda_{\text{PF}}$,
\begin{equation}
  \lim_{t \to \infty} \frac{\dot{x}_i}{x_i} = \lambda_{\text{PF}}
  \label{eq:convergence_growth_rate}
\end{equation}
If the graph is disconnected, the agents in disconnected components $C_i$
have uncoupled equations of the form (\ref{eq:knowledge-growth}) that can
be solved separately. Let $G=(V,E)$ be a graph with connected components
$C_1,C_2,...,C_l$. The set of eigenvalues of $G$, i.e. the spectrum of
$G$, is the union of sets of eigenvalues of the components. Thus,
$\lambda_{\text{PF}}(G)=\max_{j} \{ \lambda_{\text{PF}}(C_j) \}$
\citep{robinson80:_digrap, chung97:_spect_graph_theor}.

In the following, we repeat here the Perron-Frobenius theorem in a
formulation convenient to our context
\citet{seneta06:_non_matric_and_markov_chain}.
\begin{theorem}[The Perron-Frobenius Theorem]
  Let $A$ be a non-negative matrix. Then (1) the spectral radius is an
  eigenvalue, (called $\lambda_{\text{PF}}$) and all other eigenvalues
  are smaller or equal in absolute value; (2) $\lambda_{\text{PF}}$ is
  associated to one or more non-negative eigenvectors and, (3)
  $\lambda_{\text{PF}}$ is bounded from below and above as follows:
  $\min_i \sum_{j} a_{ij} \le \lambda_{\text{PF}} \le \max_i \sum_{j}
  a_{ij}$.

  If, in addition, $A$ is an irreducible matrix, then (4)
  $\lambda_{\text{PF}}$ has multiplicity 1 and (5) the associated
  eigenvector is positive.

  If, in addition, $A$ is a primitive matrix, then (6)
  $\lambda_{\text{PF}}$ is strictly greater in absolute value than all
  other eigenvalues.
\label{theor:Perron-Frobenius}
\end{theorem}

Notice that, going from non-negative to irreducible matrices the
eigenspace of $\lambda_{\text{PF}}$ reduces from several non-negative
eigenvectors to only one positive eigenvector.  In the limit of large
$t$, the terms related to the largest real eigenvalues will dominate in
Eq. (\ref{eq:general_solution}). In particular, it can be shown that for
large $t$, the vector $x(t)$ converges (in direction) to a linear
combination of Perron eigenvectors (associated to the Perron eigenvalue
$\lambda_{\text{PF}}$) \citep{horn90:_matrix_analy}, where the specific
linear combination may depend on the initial conditions. In particular,
if the adjacency matrix is primitive, the Perron eigenvector is unique
and there is a unique stable attractor.  Interpreting the result in our
model, one can say that
\begin{prop}
  If the graph of interaction between agents is primitive, there is a
  unique asymptotic distribution of relative values of knowledge
  $\mathbf{x} / \sum_{j=1}^n x_j$ given by the Perron eigenvector of the
  adjacency matrix $\mathbf{A}$.
\label{prop:unique_knowledge_distribution}
\end{prop}
If the assumption of primitivity of the matrix falls, in particular if
the matrix is non-negative but not irreducible, then there are, in
general, several Perron eigenvectors and thus several possible equilibria
for the relative values of knowledge, depending on the initial condition.

It is useful to look at an alternative but equivalent way to characterize
a primitive graph. A graph $G$ is primitive if and only if it is
connected and the greatest common divisor of the set of length of all
cycles in $G$ is $1$ \citep{xu03:_theor_applic_graph}. This means, for
instance, that the connected graph consisting of two connected nodes is
not primitive as the only cycle has lenght $2$ (since the link is
undirected a walk can go forward and backward along the link). Similarly,
a chain or a tree is also not primitive, since all cycles have only even
length. However, if we add one link in order to form a triangle, the
graph becomes primitive. The same is true, if we add links in order to
form any cycle of odd length. We can state the following result.
\begin{prop}
  If the graph $G$ is connected, the presence of one cycle of odd lenght
  is a sufficient condition for the primitivity of $G$ and hence for the
  uniqueness of the relative knowledge distribution $\mathbf{x} /
  \sum_{j=1}^n x_j$ given by the Perron eigenvector.
  \label{prop:unique_knowledge_distribution_odd_cycle}
\end{prop}

We now discuss the relation between walks in the graph and growth rate of
knowledge. In our model, a walk in the graph corresponds to a sequence of
agents contributing to their individual knowledge to their neighbors in
the walk in order to generate a sequence of recombined knowledge. As
mentioned in the beginning, each component of the power $k$ of the
adjacency matrix, $\left( \mathbf{A}^k \right)_{ij}$, gives the number of
walks of length $k$ from node $i$ to node $j$. Considering the vector
$\mathbf{u}=(1,...,1)$, we have that $n_k:=\mathbf{u}^T \mathbf{A}^k
\mathbf{u}$ is the number of all walks of length $k$ among all nodes in
$G$. Since the adjacency matrix is symmetric we have that
$\mathbf{u}=\sum a_i \mathbf{w}_i$ where $\mathbf{w}_i$ is the
eigenvector of $\mathbf{A}$ associated with the eigenvalue $\lambda_i$.
It follows that $n_k = \sum_i |a_i|^2 \lambda_i^k$.  For large $k$, we
have approximately $n_k \sim \lambda_{\text{PF}}^k$
\citep{chung07:_compl_graph_networ}, and we get
\begin{equation}
  \frac{n_k - n_{k-1}}{n_{k-1}} \sim \lambda_{\text{PF}}-1
\end{equation}
Thus, the largest real eigenvalue $\lambda_{\text{PF}}$ of the graph
measures the growth rate of the number of walks of length $k$ when the
length increases by one.
as well as the growth factor of the number of knowledge recombinations in
the network of collaborations. As we have seen in the first part of this
section, $\lambda_{\text{PF}}$ coincides also with the asymptotic growth
rate of knowledge in time.  Therefore, the faster the number of walks in
the graph (and thus of knowledge recombinations) grows with the length of
the walks, the faster also grows in time the knowledge of the agents
involved.  One should not confuse the two growth rates, one in time and
the other with respect to walk length (which does not vary in time, as we
are analyzing a static network).

A similar interpretation comes from the Rayleigh-Ritz theorem
\citep{horn90:_matrix_analy} which states that:
\begin{equation}
  \lambda_{\text{PF}} = \max_{\mathbf{x} \ne 0} 
  \frac{\mathbf{x}^T\mathbf{A}\mathbf{x}}{\mathbf{x}^T\mathbf{x}} 
\label{eq:rayleigh_ritz}
\end{equation}
where the maximum is obtained for the Perron eigenvector associated with
$\lambda_{\text{PF}}$. Here, $\mathbf{x}$ can be any vector in
$\mathbb{R}^n$. $x_ix_j$ can be interpreted as the result of the
recombination of the knowledge of agents $i$ and $j$ if they are
connected. Accordingly, one can interpret the right-hand side of Eq.
(\ref{eq:rayleigh_ritz}) as the maximum number of total knowledge
recombinations, $\mathbf{x}^T\mathbf{A}\mathbf{x}=\sum_{i,j} x_i a_{ij}
x_j$, normalized to the absolute total knowledge,
$\mathbf{x}^T\mathbf{x}=\sum_{i=1}^n x_i^2$. Some other results relate
$\lambda_{\text{PF}}$ to the number of links or the degree of the nodes
in the graph. For instance, the Perron-Frobenius theorem states that
$\lambda_{\text{PF}}$ is bounded from below and above by the minimum and
maximum degree respectively ($d_i=\sum_j a_{ij}$ is the degree of node
$i$). This means, that the higher (minimum or maximum) the degree of the
nodes in the graph, the higher $\lambda_{\text{PF}}$ and thus the
knowledge growth rate. We denote the maximum degree in $G$ by $\Delta$.
Then, a better lower bound holds so that $\sqrt{\Delta} \le
\lambda_{\text{PF}}(G) \le \Delta$. We refer to
\citet{cvetkovic90:_larges_eigen_graph, cvetkovic95:_spect_graph} for
other inequalities involving $\lambda_{\text{PF}}$.

There is also a result about the inequality of the growth rate of
knowledge across agents. For a primitive matrix $\mathbf{A}$ one can show
\citep{boyd:_linear_dynam_system} that the Perron-Frobenius eigenvector
associated with the eigenvalue $\lambda_{\text{PF}}$ is the solution to
the following optimization problem
\begin{equation}
    \max_{\mathbf{x}>0} \min_{1\le i \le n} \frac{\sum_{j=1}^n a_{ij} x_j}{x_i} 
\end{equation}
where $\sum_{j=1}^n a_{ij} x_j = \left( \mathbf{A} \mathbf{x} \right)_i =
\dot{x}_i$. The Perron eigenvector is the vector that maximizes the
minimum growth factor over all agents $i$ and also minimizes the maximum
growth factor. By maximizing the minimum growth factor we obtain balanced
growth \citep{boyd:_linear_dynam_system}. In terms of our model:
\begin{prop}
  If the graph $G$ is primitive, the unique stable distribution of
  relative knowledge values $\mathbf{x} / \sum_{j=1}^n x_j$ to which the
  dynamics (\ref{eq:knowledge-growth}) converges, is also the
  distribution that minimizes the difference between maximum and minimum
  growth rates across agents.
\label{prop:min_max_knowledge_rate}
\end{prop}

From the results above we can conclude that the utility function of agent
$i$ in Eq. (\ref{eq:utility}) increases with the number of walks in the
connected component to which agent $i$ belongs to. On the other hand the
utility decrease with the degree of the agent. Therefore it is best for
an agent to be able to reach the other agents through many walks but to
have not too many links.  We now compare this utility function with other
similar utility functions in the literature on innovation networks that
depend on the position of an agent in the network. For instance, the
utility function of \citet{jackson96:_strat_model_social_econom_networ}
is given by
\begin{equation}
  u_i = \sum_{i=1}^n \delta^{d(i,j)} - c d_i
  \label{eq:jackson-utility}
\end{equation}
where $0 \le \delta \le 1$ and $d(i,j)$ is the length of the shortest
path from node $i$ to node $j$. Other examples are those introduced by
\citet{holme06:_dynam_networ_agent_compet_high,
  carvalho07:_socioec_networ_long_range_inter} and \citet{fabrikant03,
  corbo05}.

The cost term in our utility function (\ref{eq:utility}) is the same as
in (\ref{eq:jackson-utility}).  The difference is in the benefit term:
while the latter utility function only considers the shortest path we
take into account all walks across. It has been argued that that
knowledge gets transferred not only along the shortest path but also
along all other paths in the network
\citep{wasserman94:_social_networ_analy}. Accordingly, all agents to
which agent $i$ is indirectly connected to, contribute to the utility of
agent $i$ in our model.
\citet{goyal00:_Noncooperative_model_network_formation, kima07:_networ}
introduce a utility function of the form
\begin{equation}
  u_i = |C_i| - c d_i
  \label{eq:goyal-utility}
\end{equation}
where $|C_i|$ is the size of the connected component of agent $i \in
C_i$, that is the number of agents who can be reached by agent $i$ in the
network $G$. This utility function takes into account all agents that
agent $i$ can reach and it is higher the more agents there are in its
connected component. The difference between (\ref{eq:goyal-utility}) and
our utility function (\ref{eq:utility}) is that, in our model, not only
the number of agents that can be reached (size of the connected component
$|C_i|$) but also the structure of the component contributes to the
utility of the agent.

\section{Efficiency}
\label{sec:efficiency}

In this section we define the efficiency of the system (the social
optimum for all agents) and we show that if cost is not too high
($c<1/2$), the complete graph is efficient. In the following section
(\ref{sec:stability}) we show that however, during the evolution, the
system does not necessarily reach the efficient network and can very well
stabilize in inefficient networks. For the investigation of the set of
efficient graphs in the whole range of cost, see
\citep{koenig07:_effic_stabil_dynam_innov_networ}
\begin{defn}
  The performance $\Pi(G)$ of the network $G$ is defined as the sum of
  the individual utility
  \begin{equation}
    \begin{array}{ll}
      \Pi(G) & = \sum_{i=1}^n u_i \\
      & = \sum_{i=1}^n \left( \lambda_{\text{PF}}(C_i) - c d_i \right) \\
      & =  \sum_{i=1}^n \lambda_{\text{PF}}(C_i) - 2 m c\\
    \end{array}
    \label{eq:total_returns}
  \end{equation}
  where $m$ denotes the number of edges in $G$ and $C_i$ is the connected
  component to which agent $i$ belongs.
\end{defn}

If $G$ is connected, then there is obviously only one component. The idea
of definition (\ref{eq:total_returns}) is that, in order to maximize the
performance of the system, one has to maximize total knowledge growth
while minimizing the total cost.  $\Pi$ is given by the sum of the
individual asymptotic net returns, which is just the sum of the
asymptotic individual knowledge growth rates $\frac{\dot{x_i}}{x_i}$
minus the total cost for all links.

The network $G^*$ is called \textit{efficient}, if it maximizes $\Pi$ over
the set of all possible graphs with a given number of nodes:
\begin{equation}
  G^* = \argmax_{G}\{\Pi(G) : |V(G)|=n \}
  \label{def:efficent_graph}
\end{equation}

Following \citet{cvetkovic90:_larges_eigen_graph} , we will denote the
star with $n$ nodes (and $n-1$ edges) as $K_{1,n-1}$ and the complete
graph with $n$ nodes as $K_n$.

We can immediately determine the efficient network, in the special case
of null costs, $c=0$. The case $c<1/2$ requires some more work.
\begin{prop}
  If costs are zero, $c=0$, then the complete graph $K_n$ is the
  efficient graph. Its performance is given by
  $\Pi(K_n)=n\lambda_{\text{PF}}(K_n)=n(n-1)$.
\end{prop}
\begin{proof}
  If costs are zero, then total asymptotic net returns are $\Pi = n
  \lambda_{\text{PF}}$. The graph with the highest eigenvalue is the
  complete graph $K_n$ with $\lambda_{\text{PF}}(K_n)=n-1$
  \citep{hong93:_bound}. $\Box$
\end{proof}
\begin{prop} The complete graph $K_n$ is efficient for $c <
  \frac{1}{2}$. For costs $c \ge n$ the empty graph is efficient.
  \label{prop:complete_empty_graph_efficiency}
\end{prop}
\begin{proof} 
  Since for the complete graph it is $\lambda_{\text{PF}}=n-1$ and
  $m=\frac{n(n-1)}{2}$, its performance is $\Pi(K_n)=
  n(n-1)-2\frac{n(n-1)}{2}c=n(n-1)(1-c)$.

On the other hand, the largest real eigenvalue $\lambda_{\text{PF}}$ of a
graph $G$ with $m$ edges is bounded from above so that
$\lambda_{\text{PF}} \le \frac{1}{2}(\sqrt{8m+1}-1)$
\citep{stanley87:_bound_spect_radius_graph_edges}. For the performance of
the system we then have
\begin{equation}
    \label{eq:growth_upper_bound}
    \Pi = \sum_{i=1}^n \lambda_{\text{PF}}(G_i) - 2mc \le 
    n \max_{1 \le i \le n} \lambda_{\text{PF}}(G_i) - 2mc  \le  
    \frac{n}{2}(\sqrt{8m+1}-1) - 2cm := b(n,m,c)
\end{equation}
with $n \le m \le {n \choose 2}$. For fixed cost $c$ and number of nodes
$n$, the number of edges maximizing Eq. (\ref{eq:growth_upper_bound}) is
given by $m^*=\frac{n^2-c^2}{8c^2}$ if $\frac{n^2-c^2}{8c^2} < {n \choose
  2}$ and $m^*=\frac{n(n-1)}{2}$ if $\frac{n^2-c^2}{8c^2} > {n \choose
  2}$.  The graph with the latter number of edges is the complete graph.
Inserting $m^*$ into Eq. (\ref{eq:growth_upper_bound}) yields

  \begin{equation}
    \label{eq:growth_upper_bound_m*}
    b(n,m^*,c) = 
    \begin{cases} 
      \frac{n}{2}(\sqrt{\frac{n^2-c^2}{c^2}+1}-1)-\frac{n^2-c^2}{4c} &
      c>\frac{n}{2n-1} \\
      n(n-1)(1-c)=\Pi (K_n) &
      c<\frac{n}{2n-1} \\
    \end{cases}
  \end{equation}

  \begin{figure}[htpb]
    \psfrag{b(n,m*,c)}[c][][3][0]{$b(n,m^*,c)$} 
    \psfrag{Kn}[c][][3][0]{$\leftarrow K_n$} 
    \psfrag{c}[c][][3][0]{$c$}
    \centering \centerline{\scalebox{0.4}{\includegraphics[angle=0]
        {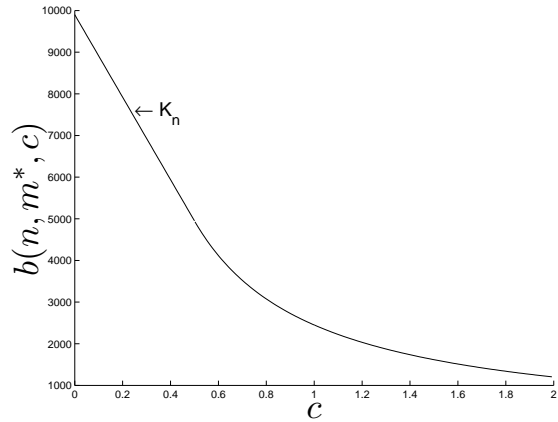}}}
    \caption{Upper bound $b(n,m^*,c)$ of Eq.
      (\ref{eq:growth_upper_bound_m*}) for $n=100$ and varying costs $c$.
      For $c \le \frac{n}{2n-1}$ the upper bound is given by the complete
      graph $K_n$.}
    \label{fig:upper bound}
  \end{figure}
  The bound for $c \le \frac{n}{2n-1} \sim \frac{1}{2}$ coincides with
  the performance of the complete graph, $K_n$ which is therefore the
  efficient graph. If instead $c=n$ then $m^*=0$ and the efficient graph
  is the empty graph. This concludes the proof. Notice that a similar
  result can be obtained using an alternative bound for connected graphs,
  $\lambda_{\text{PF}} \le \sqrt{2m-n+1}$ due to \cite{hong93:_bound}.
  $\Box$
\end{proof}

\section{Network Evolution}
\label{sec:network_evolution}

In our model we assume that agents have a certain inertia for creating
new links and evaluating their existing ones. The rate at which links are
formed is much slower than the rate at which knowledge flows (and the
knowledge stocks of agents change). In other words, there are two
different time scales in our dynamical system: the fast dynamics of the
knowledge levels and returns and the slow evolution of the network. The
returns immediately reach their quasi-equilibrium state, whereas the
network remains unchanged during this short adaptation time.  One can say
that the variables with the fast dynamics are ``slaved'' by the variables
with the slow dynamics \citep{haken77:_syner_introd} (see
\citet{gross07:_adapt_coevol_networ_review} for a review)\footnote{This
  principle has been used e.g by
  \citet{jain98:_autoc_sets_growt_compl_evolut, jain01} in the context of
  evolutionary biology. Subsequently
  \citet{saurabh07:_growt_knowl_compl_evolv_networ} have applied their
  model to an innovation system.}. We assume that the knowledge growth
rate has got very close to its asymptotic value when agents create new
links.

In the network evolution process, pairs of agents are asynchronously
updated. Let $T$ count the number of such updates. At every network
update $T$ the following steps are taken: (1) Two agents, not already
connected, are uniformly selected at random to form a link and the
creation time $T$ (birth date) is recorded for that link. (2) All links
that have been previously created and that are as old as $\tau$ are
evaluated (with $\tau$ as an exogenous parameter).  For the evaluation of
a link, the incident agents compare their current utility at $T$ with the
utility before the creation at $T - (\tau+1)$, i.e. before the birth date
of the link. The link is maintained only if both agents strictly increase
their utility\footnote{Note that in the mean time the network and also
  the neighbors of the agents that are evaluating the link may have
  different knowledge values which in turn affects the utility of the
  agents. This means that the utility of the agents may have increased
  due to different reasons than the link that is currently evaluated. In
  our model agents do not distinguish which of their links is responsible
  for an increase or decrease in their utility separately but they rather
  observe the overall effect on their utility by all their links at the
  same time.}, otherwise the link is removed. (3) Finally. the age of all
links is increased by one, $T \rightarrow T+1$. This process is
represented in the following algorithm.

\begin{itemize}

\item[\textbf{1}] Initialization: empty graph

\item[\textbf{2}] \textbf{quasi-equilibrium} (fast knowledge
  growth/decline):
  
With $\mathbf{A}$ fixed, the knowledge grows according to Eq.
  (\ref{eq:knowledge-growth}) to reach constant growth rates
  (``balanced'' growth).

\item[\textbf{3}] \textbf{perturbation}: network update (slow network
  evolution)
  
\begin{itemize}
  \item[(i)] A pair of agents is randomly chosen to create a link.
  \item[(ii)] The performance of the links attaining an age of $\tau$ is
    evaluated\footnote{If $\tau=1$ the link that is created is
      immediately evaluated afterwards.}:
    
    \begin{tabular}{lll}
      if both utilities have increased  & $\rightarrow$ &  keep the link\\
      otherwise & $\rightarrow$ & remove the link \\
    \end{tabular}
   
\end{itemize}

\item[\textbf{4}] Stop the evolution if the network is
  stable\footnote{The notion of stability is defined
    in (\ref{def:stability})}, otherwise $T \rightarrow T+1$ and go to
  \textbf{2}

\end{itemize}

\begin{figure}[htpb]
  \begin{center}
    \begin{pspicture}(0,1)(6,4) 
      \rput(1,3.5){\rnode{A}{\psframebox{
            \begin{tabular}{c}
              \textbf{initialization}
            \end{tabular}
          }}} \rput(1,2){\rnode{B}{\psframebox{
            \begin{tabular}{c}
              $x_i$ reach \\ 
              \textbf{quasi-equilibrium}
            \end{tabular}
          }}} \rput(5,2){\rnode{C}{\psframebox{
            \begin{tabular}{c}
              \textbf{perturbation}\\
              of $ a_{ij}$
            \end{tabular}
          }}} \ncline[linewidth=0.5pt,linestyle=dotted]{->}{A}{B}
      \ncarc[linewidth=1pt,arcangle=40]{->}{B}{C}
      \ncarc[linewidth=1pt,arcangle=40]{->}{C}{B}
    \end{pspicture}
  \end{center}
\end{figure}

In the version of the model analyzed here the links that pass the
evaluation $\tau$ network updates after their creation remain in the
graph forever. Some of the results presented here still hold if this
hypothesis is relaxed and links are evaluated again in the future and
possibly deleted, but we do not consider this case in the present paper.
The local process of formation and deletion of links described above
intends to mimic the process by which selfish agents improve their
utility through a trial and error method.

\section{Stability}
\label{sec:stability}

In this section we first give a definition of stability as a stationary
network resulting from the network formation process described in the
previous section. We then give two examples of stable networks that are
not efficient, the star $K_{1,n-1}$ and the clique $K_n$. Finally we
derive an upper bound for the cost of links above which the complete
graph is not reachable (and for costs $c<\frac{1}{2}$ it is the efficient
network). In the proofs of this section we assume an evaluation period of
$\tau=1$. This means that links are evaluated immediatley after they are
created.

The network evolution is in a stable equilibrium if it is bilaterally
stable. A network $G$ is bilaterally stable if and only if no pair of
agents, $i$ and $j$, can create a bilateral connection such that the
utilities of both agents at the evaluation period $\tau$ are higher than
the current utilities. More formally,
\begin{defn}
  For a fixed value of $\tau$ $G$ is bilaterally stable at time $t$ if
  there does not exist a pair $i,j$ such that both $u_i(t+\tau)>u_i(t)$
  and $u_j(t+\tau)>u_j(t)$.
  \label{def:stability}
\end{defn}

This definition is similar to the notion of pairwise stability introduced
earlier by
\citet{jackson03:_survey_models_network_formation_stability_efficiency}.
Let $G+uv$ denote the graph obtained by adding a link $uv$ to the
existing graph $G$. The addition of one edge $uv$ leads to a change in
the individual utility $u_i$ (cf. Eq. (\ref{eq:utility}))
\begin{equation}
  \begin{array}{ll}
    \Delta u_i & = u_i(G+uv) - u_i(G) \\
    & = \lambda_{\text{PF}}(G+uv) -
    c (d_i + 1) - ( \lambda_{\text{PF}}(G) - c d_i) \\
    & = \Delta \lambda_{\text{PF}} -c
  \end{array}
\end{equation}

\begin{prop}
  For $\tau=1$ and any value of cost $c$, there exists a number of nodes
  $n$ such that the star $K_{1,n-1}$ is bilaterally stable .
  \label{prop:star_stability}
\end{prop}
\begin{proof} 
  We consider a graph with consisting in the star $K_{1,n-1}$ as a
  subgraph and some isolated nodes and we show that if $n$ is large
  enough, the benefit of any additional link is smaller than a given cost
  $c$. As shown in Fig. \ref{fig:K_1_8}, there are only three types of
  links that can be added, either between the nodes of the star or by
  attaching a disconnected node.  \begin{figure}
  \begin{center}
    \begin{pspicture}(-2,-2)(2,2) 
      
      \psset{nodesep=1pt}
 
      \cnodeput(1.0000,0.0000){E2}{10} 

      \cnodeput(-3,0){A1}{1} \cnodeput(-1.5858,1.4142){B1}{2}
      \cnodeput(-3.0000,2.0000){C1}{3} \cnodeput(-4.4142,1.4142){D1}{4}
      \cnodeput(-5.0000,0.0000){E1}{5} \cnodeput(-4.4142,-1.4142){F1}{6}
      \cnodeput(-3.0000,-2.0000){G1}{7} \cnodeput(-1.5858,-1.4142){H1}{8}
      \cnodeput(-1.0000,-0.0000){I1}{9}
      
      \ncline[linestyle=dotted]{I1}{E2}\ncput*{(i)}
      \ncline[linestyle=dotted]{C1}{F1}\ncput*{(ii)}
      \ncarc[linestyle=dotted,arcangle=40]{-}{A1}{E2}\ncput*{(iii)}

      \ncline{A1}{B1} \ncline{A1}{C1} \ncline{A1}{D1} \ncline{A1}{E1}
      \ncline{A1}{F1} \ncline{A1}{G1} \ncline{A1}{H1} \ncline{A1}{I1}

%
%
%
%
%
%
%
%

      \cnodeput[fillstyle=solid](-3,0){A1}{1}

    \end{pspicture}
  \end{center}
  \caption{A star $K_{1,8}$ and an additional edge, appended either from an isolated node to the nodes in the star or between two nodes in the star.}
  \label{fig:K_1_8}
\end{figure}
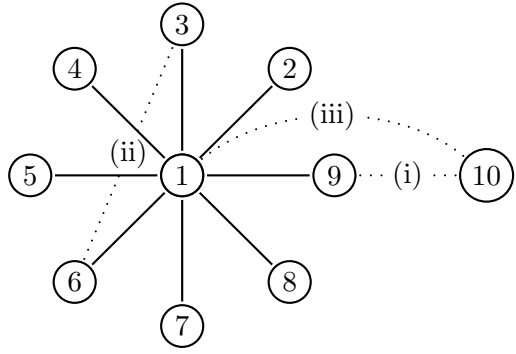 

\begin{itemize}  
\item By adding a leaf to the central node of the star $K_{1,n-1}$, (link
  (iii) in Fig. \ref{fig:K_1_8}), the change in eigenvalue is $\Delta
  \lambda_{\text{PF}} = \sqrt{n}-\sqrt{n-1}$ which is monotonically
  decreasing with $n>0$. For any given $c$ if $\sqrt{n}-\sqrt{n-1} < c$
  then no new node will be attached to $K_{1,n-1}$.
  
  \item If a link is created from a disconnected node to a peripheral
    node in the star (link (i) in Fig. \ref{fig:K_1_8}), then the resulting
    eigenvalue is smaller than the eigenvalue obtained from the link
    created to the central node in the star. This comes from the fact
    that the former adjacency matrix is not stepwise while the latter is,
    see \citet{brualdi86:_spect_radius_connec_graph}.
  
  \item The case of a link between two nodes in the
    star  (link (i) in Fig. \ref{fig:K_1_8}) requires a little more work.
    We use an upper bound due to \citet{maas87:_pertur} on the increase
    of the largest real eigenvalue $\Delta \lambda_{\text{PF}}$ of a
    connected undirected graph $G$, if an edge $ij$ is added.  The upper
    bound depends on the component $i$ and $j$ of the eigenvector
    $\mathbf{x}$ associated to $\lambda_{\text{PF}}$:
    \begin{equation}
      \label{eq:maas}
      \lambda_{\text{PF}}(G+ij) - \lambda_{\text{PF}}(G) 
      < 1+\delta - \frac{\delta(1+\delta)(2+\delta)}{(x_i+x_j)^2+\delta(2+\delta+2x_ix_j)}
    \end{equation}
    where $\delta$ denotes the minimum degree in the graph $G$. With this
    upper bound we can compute the change in individual utility by
    the addition of an edge, $\Delta u_i=\Delta \lambda_{\text{PF}} - c$.

    The characteristic polynomial of the star $K_{1,n-1}$ is given by
    $(\lambda^2-(n-1))\lambda^{n-2}$. Thus the largest real eigenvalue is
    $\lambda_{\text{PF}}=\sqrt{n-1}$. The corresponding normalized
    eigenvector is given by
    $\frac{1}{2(n-1)}(1,...,1,\sqrt{n-1},1,...,1)^T$. Applying Eq.
    (\ref{eq:maas}) to the star $K_{1,n-1}$ gives $ \Delta
    \lambda_{\text{PF}} = \left( \lambda_{\text{PF}}(K_{1,n-1}+ij) -
      \lambda_{\text{PF}}(K_{1,n-1}) \right) < 2 - \frac{4n^2}{1+2n^2}$
    which leads to $\Delta u_i < \frac{2}{1+2n^2}-c$. For $n \ge
    \sqrt{\frac{2-c}{2c}}$, $\Delta u_i$ becomes negative and therefore
    adding the link is not profitable.
 \end{itemize}   
 Combining the results above, and noticing that $ \sqrt{n}-\sqrt{n-1} >
 \frac{2}{2n^2+1}$, we can conclude that, for any $c$, no link of either one of the
 three types is added to the star for $n$ large enough. $\Box$
\end{proof}

Notice that the bound of \citet{maas87:_pertur} which we used in the
first part of the previous proposition holds only for connected graphs
and cannot be used when adding a link that connects a graph to a
previously disconnected node. For the next proofs we will use a bound on
the increase of the largest real eigenvalue $\Delta \lambda_{\text{PF}}$
when a link is added, which depends only on the number of links and nodes
in the graph, regardless of the structure of the links.

\begin{prop}
  The change in the largest real eigenvalue, $\Delta \lambda_{\text{PF}}$
  of a graph $G$ with $m$ edges and $n$ nodes, by adding one edge to the
  graph is bounded as follows
  \begin{equation}
    \Delta \lambda_{\text{PF}} \le \frac{1}{2}( -1 + \sqrt{1+8(m+1)}  ) - \frac{2m}{n}
  \end{equation}
  \label{prop:bound_delta_lambda}
\end{prop}
\begin{proof}
  The average degree of the graph is $\bar d = \frac{2m}{n}$. A lower bound
  on the largest real eigenvalue is given by $\lambda_{\text{PF}} \ge
  \bar d$ \citep{cvetkovic90:_larges_eigen_graph}. An upper bound on the
  largest real eigenvalue is given by $\lambda_{\text{PF}} \le
  \frac{1}{2} (-1+\sqrt{1+8m})$
  \citep{stanley87:_bound_spect_radius_graph_edges}. Combining the two
  bounds yields the proposition.  $\Box$
\end{proof}

\begin{prop}
  For $\tau=1$ and any value of cost $c$, there exists a number of nodes
  $n$ such the clique $K_n$ is bilaterally stable.
  \label{prop:clique_node}
\end{prop}
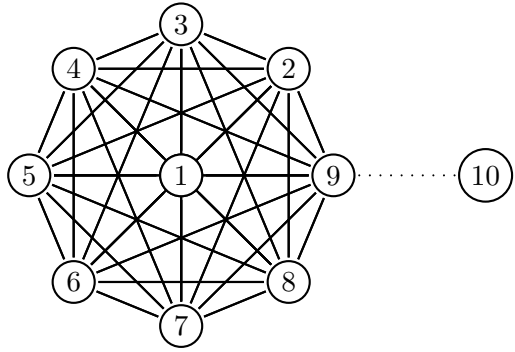
\begin{figure}
      \begin{center}
        \begin{pspicture}(-3,-2)(3,2)
      
      \psset{nodesep=1pt}
 
      \cnodeput(1.0000,0.0000){E2}{10} 

      \cnodeput(-3,0){A1}{1} \cnodeput(-1.5858,1.4142){B1}{2}
      \cnodeput(-3.0000,2.0000){C1}{3} \cnodeput(-4.4142,1.4142){D1}{4}
      \cnodeput(-5.0000,0.0000){E1}{5} \cnodeput(-4.4142,-1.4142){F1}{6}
      \cnodeput(-3.0000,-2.0000){G1}{7} \cnodeput(-1.5858,-1.4142){H1}{8}
      \cnodeput(-1.0000,-0.0000){I1}{9}
      
      \ncline[linestyle=dotted]{I1}{E2}

      \ncline{A1}{B1} \ncline{A1}{C1} \ncline{A1}{D1} \ncline{A1}{E1}
      \ncline{A1}{F1} \ncline{A1}{G1} \ncline{A1}{H1} \ncline{A1}{I1}

      \ncline{B1}{A1} \ncline{B1}{C1} \ncline{B1}{D1} \ncline{B1}{E1}
      \ncline{B1}{F1} \ncline{B1}{G1} \ncline{B1}{H1} \ncline{B1}{I1}
      
      \ncline{C1}{A1} \ncline{C1}{B1}
      
      \ncline{C1}{D1} \ncline{C1}{E1} \ncline{C1}{F1} \ncline{C1}{G1}
      \ncline{C1}{H1} \ncline{C1}{I1} \ncline{D1}{A1} \ncline{D1}{B1}
      \ncline{D1}{C1}
      
      \ncline{D1}{E1} \ncline{D1}{F1} \ncline{D1}{G1} \ncline{D1}{H1}
      \ncline{D1}{I1}

      \ncline{E1}{A1} \ncline{E1}{B1} \ncline{E1}{C1} \ncline{E1}{D1}
      \ncline{E1}{F1} \ncline{E1}{G1} \ncline{E1}{H1} \ncline{E1}{I1}

      \ncline{F1}{A1} \ncline{F1}{B1} \ncline{F1}{C1} \ncline{F1}{D1}
      \ncline{F1}{E1}
       
      \ncline{F1}{G1} \ncline{F1}{H1} \ncline{F1}{I1} \ncline{G1}{A1}
      \ncline{G1}{B1} \ncline{G1}{C1} \ncline{G1}{D1} \ncline{G1}{E1}
      \ncline{G1}{F1}
      
      \ncline{G1}{H1} \ncline{G1}{I1} \ncline{H1}{A1} \ncline{H1}{B1}
      \ncline{H1}{C1} \ncline{H1}{D1} \ncline{H1}{E1} \ncline{H1}{F1}
      \ncline{H1}{G1}
      
      \ncline{H1}{I1} \ncline{I1}{A1} \ncline{I1}{B1} \ncline{I1}{C1}
      \ncline{I1}{D1} \ncline{I1}{E1} \ncline{I1}{F1} \ncline{I1}{G1}
      \ncline{I1}{H1}

      \cnodeput[fillstyle=solid](-3,0){A1}{1}

      \end{pspicture}
  \end{center}
  \caption{A complete graph $K_{9}$ and a node appended.}
  \label{fig:K_9_node}
\end{figure}
\begin{proof}
  We consider the graph $G'$ obtained by connecting a clique $K_n$ and an
  isolated node via and edge (see Fig. \ref{fig:K_9_node}). We consider
  the increase of the largest real eigenvalue, $\Delta
  \lambda_{\text{PF}}= \lambda_{\text{PF}}(G')- \lambda_{\text{PF}}(K_n)$
  and we apply the bound of prop. \ref{prop:bound_delta_lambda}. Since
  $\bar d = n-1$ in the clique, we have $\Delta \lambda_{\text{PF}} \le
  \frac{1}{2}( -1 + \sqrt{1+8(m+1)}) - (n-1)$ which is smaller than $c$
  for $n>n^*=\frac{2+c(1+c)}{2c}$, as one can check solving the
  inequality for $m=\frac{n(n-1)}{2}+1$. $\Box$
\end{proof}

There is another bound on the change of $\lambda_{\text{PF}}$ for
bilateral link deletion or creation: if the undirected connected graphs
$G$ and $G'$ differ in only one edge then
$|\lambda_{\text{PF}}(G)-\lambda_{\text{PF}}(G')| \le 1$
\cite{cvetkovic97:_eigen_graph}. This bound is weaker than the ones
previously introduced, but it is still useful to derive the following
proposition.
\begin{prop}
  If costs are higher than one, $c>1$, and $\tau=1$, then no agent will
  create any link. Any graph is bilaterally stable and in particular, the
  empty graph is bilaterally stable.
\end{prop}
Since Eq. (\ref{eq:knowledge-growth}) implicitly assumes benefit equal
$1$ from a collaboration, the case $c>1$ is somehow an extreme case,
because the cost of a link is higher than the benefit and therefore the
result above is not surprising.

\begin{figure}
  \psfrag{n}[c][][4][0]{$n$} 
  \psfrag{c}[c][][4][0]{$c^*$}
  \psfrag{K}[c][][3][0]{$\nexists K_n$}
  \centering \centerline{\scalebox{0.4}{\includegraphics[angle=0]
      {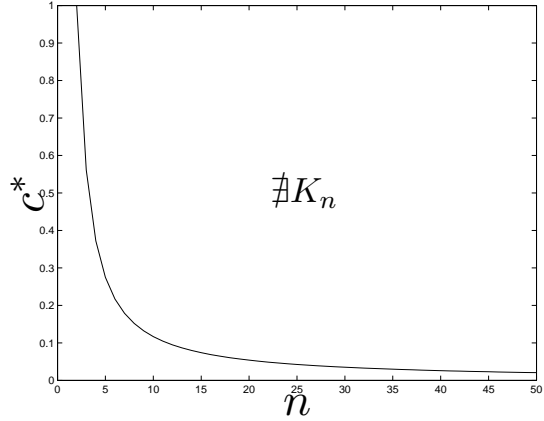}}}
  \caption{Maximal value of cost $c$ for which the complete graph can be
    obtained as an equilibrium network.}
  \label{fig:cost-upper-bound}
\end{figure}

We now prove that the efficient graph is not necessarily reached by
the evolution (see also Fig. \ref{fig:cost-upper-bound}).
\begin{prop}
  For $\tau=1$ and cost $c<1/2$ the efficient, complete graph $K_n$ of
  size $n \ge \frac{2}{c}$ cannot be reached by the network formation
  process.
 \label{prop:cost-upper-bound}
\end{prop}
\begin{proof} 
  We apply the bound of proposition (\ref{prop:bound_delta_lambda}) on
  the change in the largest real eigenvalue, $\Delta
  \lambda_{\text{PF}}$, by adding an edge to the graph $G$ with $m$
  edges. Solving the equation $\Delta \lambda_{\text{PF}}=c$ for $m$
  yields the maximal number $m^*$ of edges that can be added to a graph
  of $n$ nodes when the cost is $c$,
  $m^*(n,c)=\frac{n}{4}(-1-2c+n+\sqrt{n^2+9-2n(1+2c)})$. Notice that
  $m^*(n,c)$ decreases with increasing cost $c$. Imposing now this
  expression to be equal to one edge less than the number of edges in a
  complete graph $K_n$ of $n$ nodes, ${n \choose
    2}-1=\frac{n(n-1)}{2}-1$, we get $c^*=\frac{2}{n}$. Thus, if costs
  exceed this value then the increase in eigenvalue corresponding to the
  creation of the link that would make the graph complete, is smaller
  than the cost. Notice that $c^*$ decreases with $n$ and tends to $0$
  for large $n$, as plotted in Fig. \ref{fig:cost-upper-bound}, and
  therefore for any given $c$ there is an $n$ large enough such that the
  complete graph cannot be reached.$\Box$
\end{proof}

Similar to previous works of other authors
\cite{jackson96:_strat_model_social_econom_networ,
  goyal00:_Noncooperative_model_network_formation, corbo05} we find stars
and cliques to be stable structures for a given value of cost for the
links.  However, in this model, there is a limit size above which these
networks can be stable. Moreover, for a same level of cost, one can
obtain both a star or a complete graph (with different $n$) as stable
equilibria of the dynamics. This points to the existence of multiple
equilibria, as investigated more thoroughly in
\citep{koenig07:_effic_stabil_dynam_innov_networ}

\section{Simulation Studies of Network Evolution}
\label{sec:simulation_link_creation}

For multiple realizations (simulations) we study the evolution of the
network and the stable equilibrium networks reached by this evolution.
In order to characterize the networks, we introduce some simple network
measures:

\begin{itemize}
\item[(i)] The network density $s(G)$ of a graph $G$ is defined as the
  number of links $m$ divided by the maximum number of links
  $\frac{n(n-1)}{2}$, i.e. $s(G):=\frac{2m}{n(n-1)}$. $s(G)$ measures how
  sparse a network is.
\item[(ii)] The relative performance
  $\pi(G):=\frac{\Pi(G)}{\Pi(K_n)}=\frac{\Pi(G)}{n(n-1)(1-c)}$.  $\pi(G)$
  measures the relative performance of a network compared to the complete
  graph. In section \ref{sec:efficiency} we have shown that for costs
  $c<\frac{1}{2}$ the complete graph is efficient. Thus, a value of the
  relative performance smaller than one is a measure of the inefficiency
  of the network.
\item[(iii)] The local clustering coefficient, $C_l$, measures the
  fraction of an agent's neighbors that are also neighbors of each other.
  The global clustering coefficient, $C_g$, is the average of the local
  clustering coefficient of all agents in the network. The global
  clustering coefficient is at most one.
\end{itemize}

We first study the networks obtained with an evaluation period $\tau=1$,
which means that agents evaluate their links immediately after reaching
their balanced growth rates. We then investigate the density and
efficiency of the stable equilibrium networks that are reached, if $\tau$
is longer than $1$. This means that agents are evaluating their bilateral
links after several other agents may have created bilateral links.

\begin{figure}[htpb]
  \centering
  \begin{minipage}{.45\linewidth}
    \psfrag{relative total growth rate}[c][][4][0]{$\pi$}
    \psfrag{network updates}[c][][3][0]{$T$}
    \centerline{\scalebox{0.4}{\includegraphics[angle=0]
        {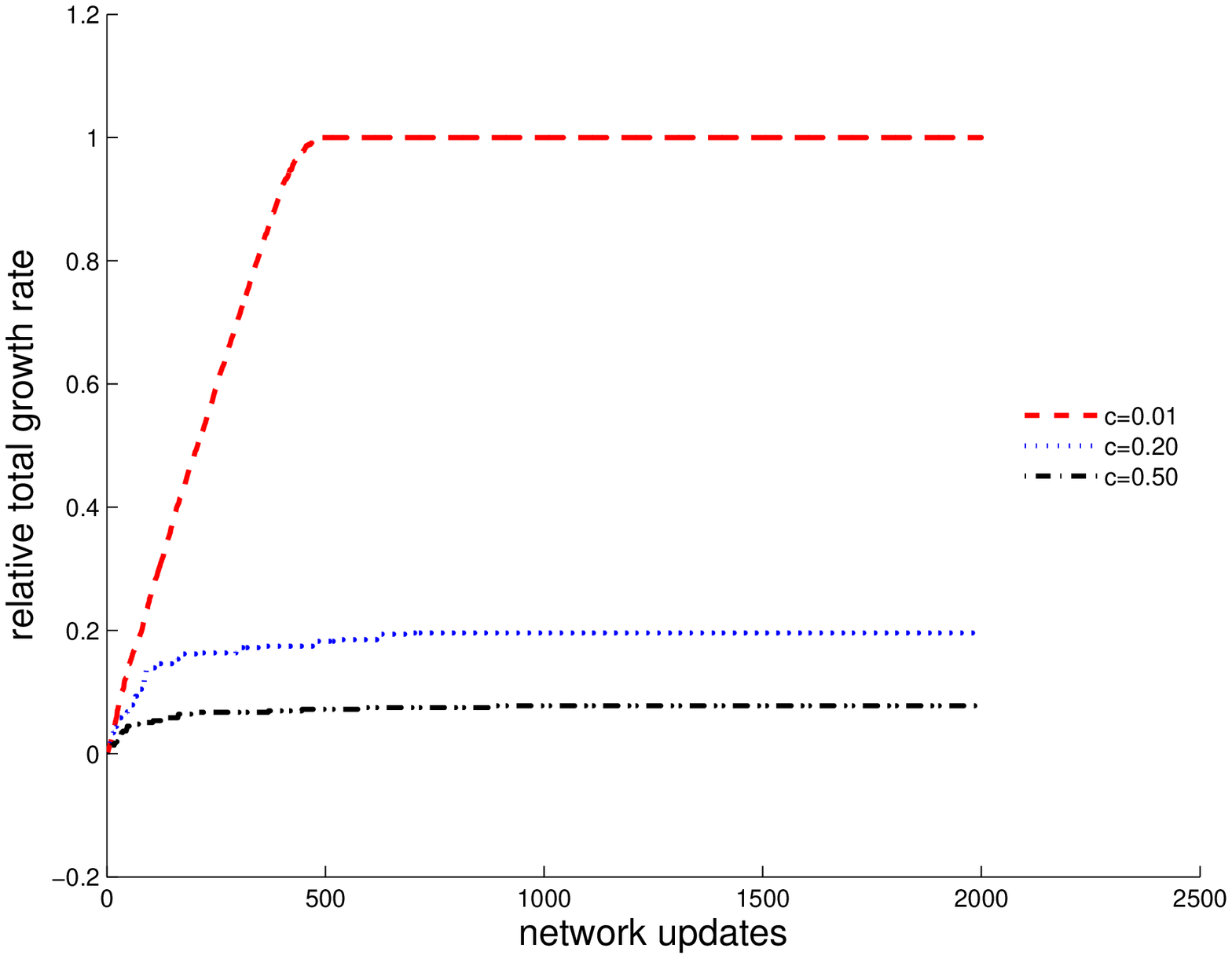}}}
  \end{minipage}
  \hfill
  \begin{minipage}{.45\linewidth}
    \psfrag{density}[c][][4][0]{$s$}
    \psfrag{network updates}[c][][3][0]{$T$}
    \centerline{\scalebox{0.4}{\includegraphics[angle=0]
        {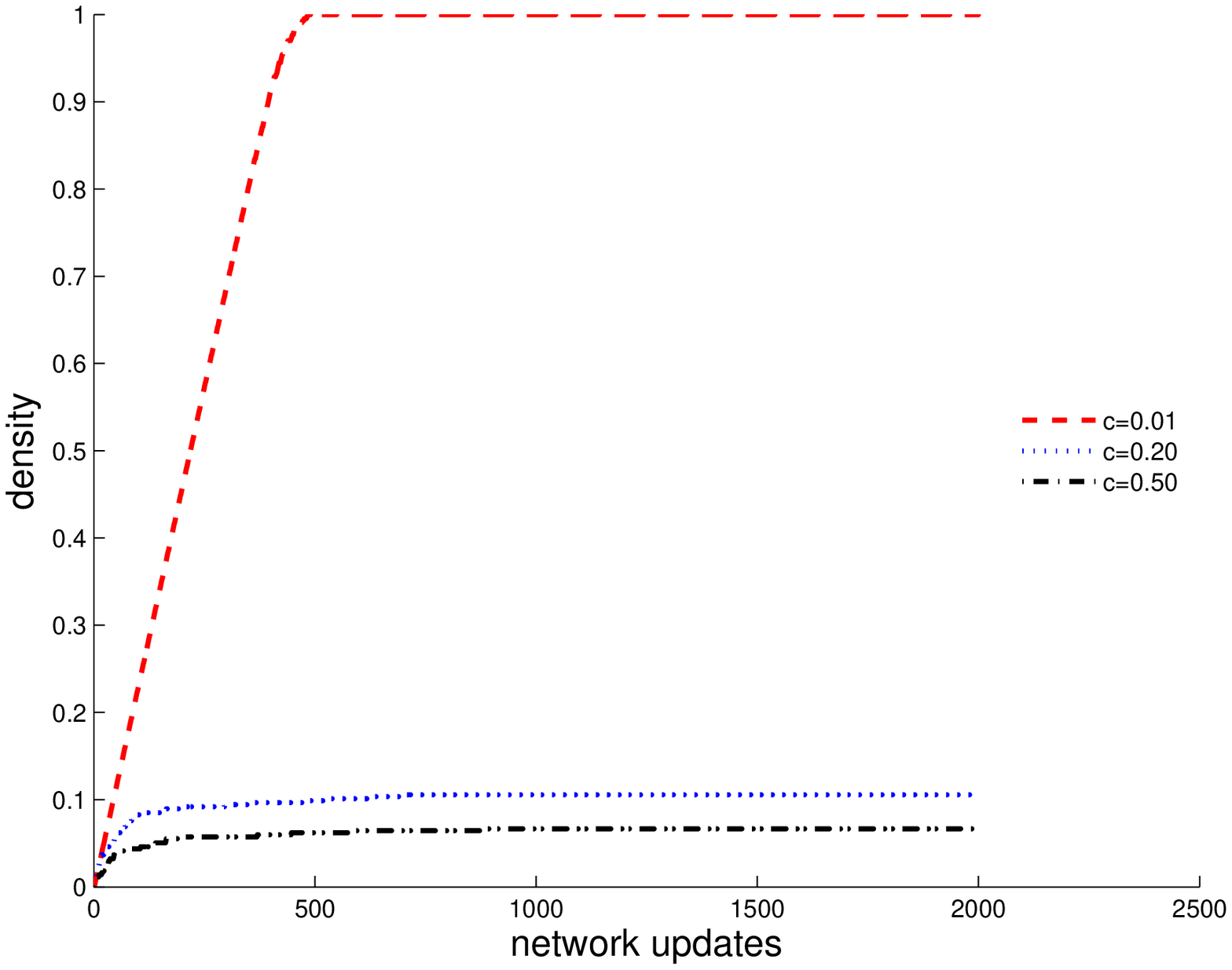}}}
  \end{minipage}
  \hfill
  \begin{minipage}{.45\linewidth}
    \psfrag{average degree}[c][][3][0]{$\langle d \rangle$}
    \psfrag{network updates}[c][][3][0]{$T$}
    \centerline{\scalebox{0.4}{\includegraphics[angle=0]
        {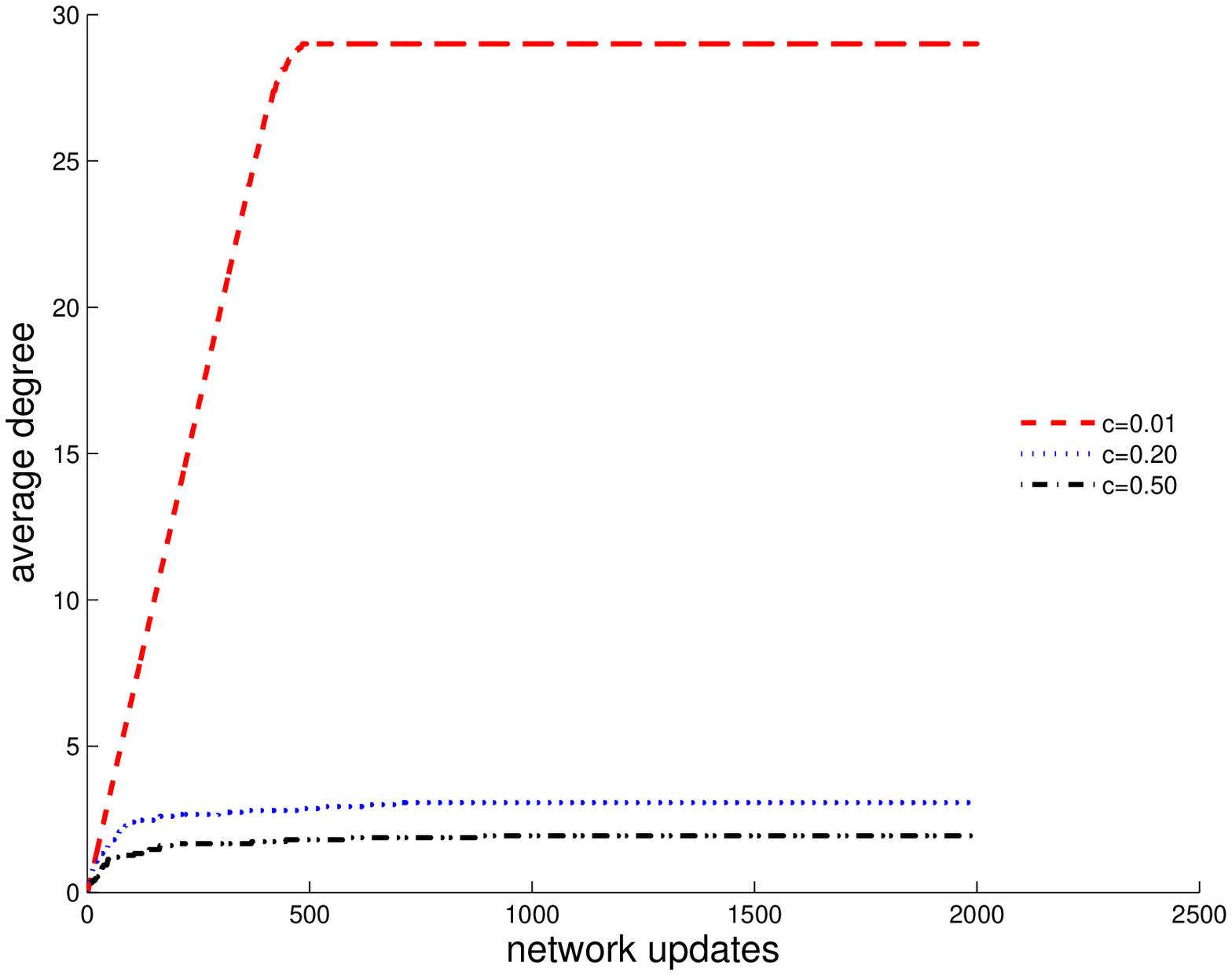}}}
  \end{minipage}
  \hfill
  \begin{minipage}{.45\linewidth}
    \psfrag{average clustering coefficient}[c][][3][0]{$C_g$}
    \psfrag{network updates}[c][][3][0]{$T$}
    \centerline{\scalebox{0.4}{\includegraphics[angle=0]
        {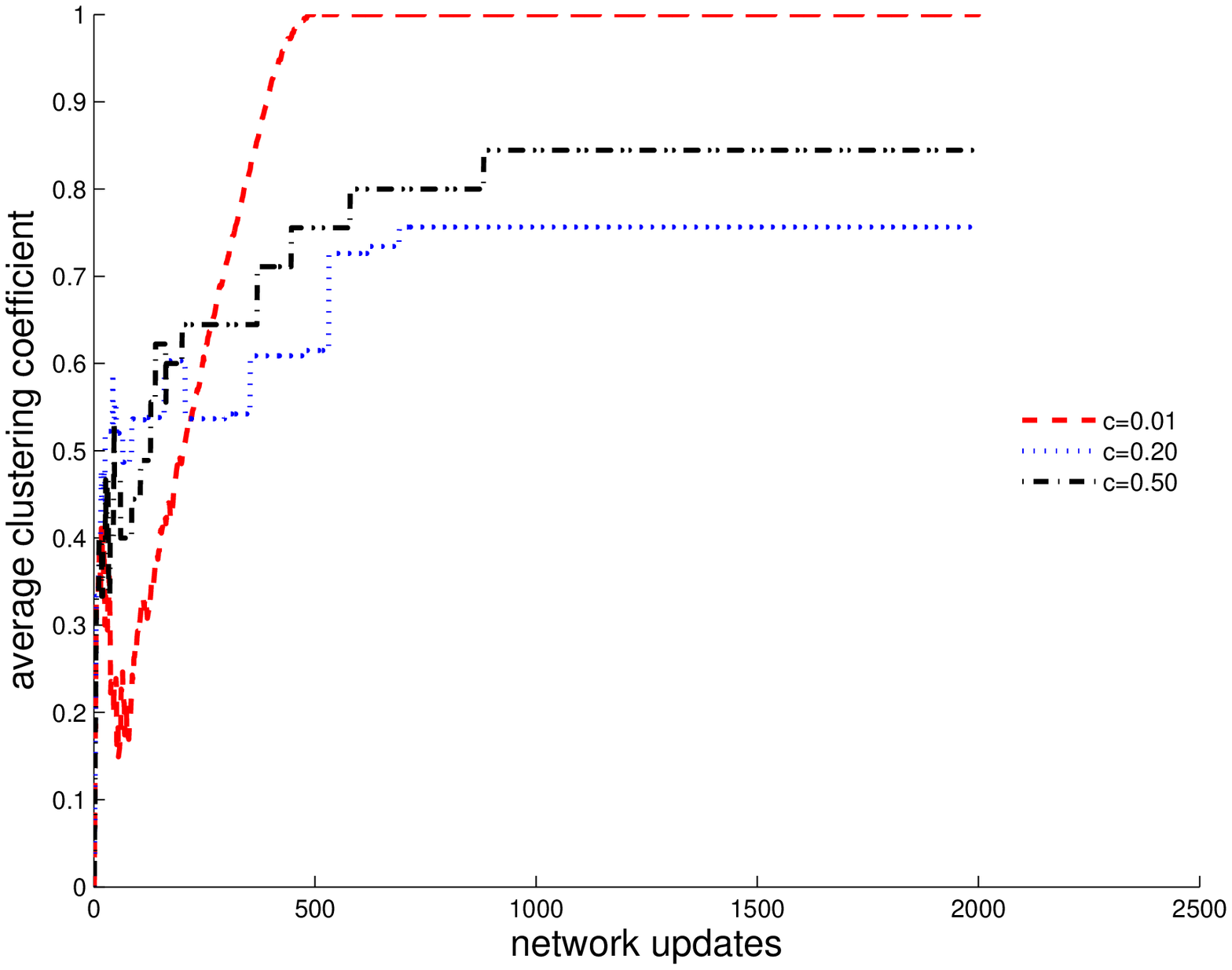}}}
  \end{minipage}
  \caption{Evolution of the network starting from an empty graph until
    reaching its stable equilibrium configuration, Fig.
    (\ref{fig:graph_plots} bottom). The stable equilibrium network for
    intermediate costs $c=0.2$ is inefficient, sparse and is highly
    clustered, while for small costs $c=0.01$ the efficient complete
    graph is realized.}
  \label{fig:graph_evolution_c_0.2}
\end{figure}

\begin{figure}
  \begin{center}
    \begin{pspicture}(0,-1)(10,16) \cnodeput*(5,14.5){A}{
        \scalebox{0.2}{\includegraphics[angle=0]
          {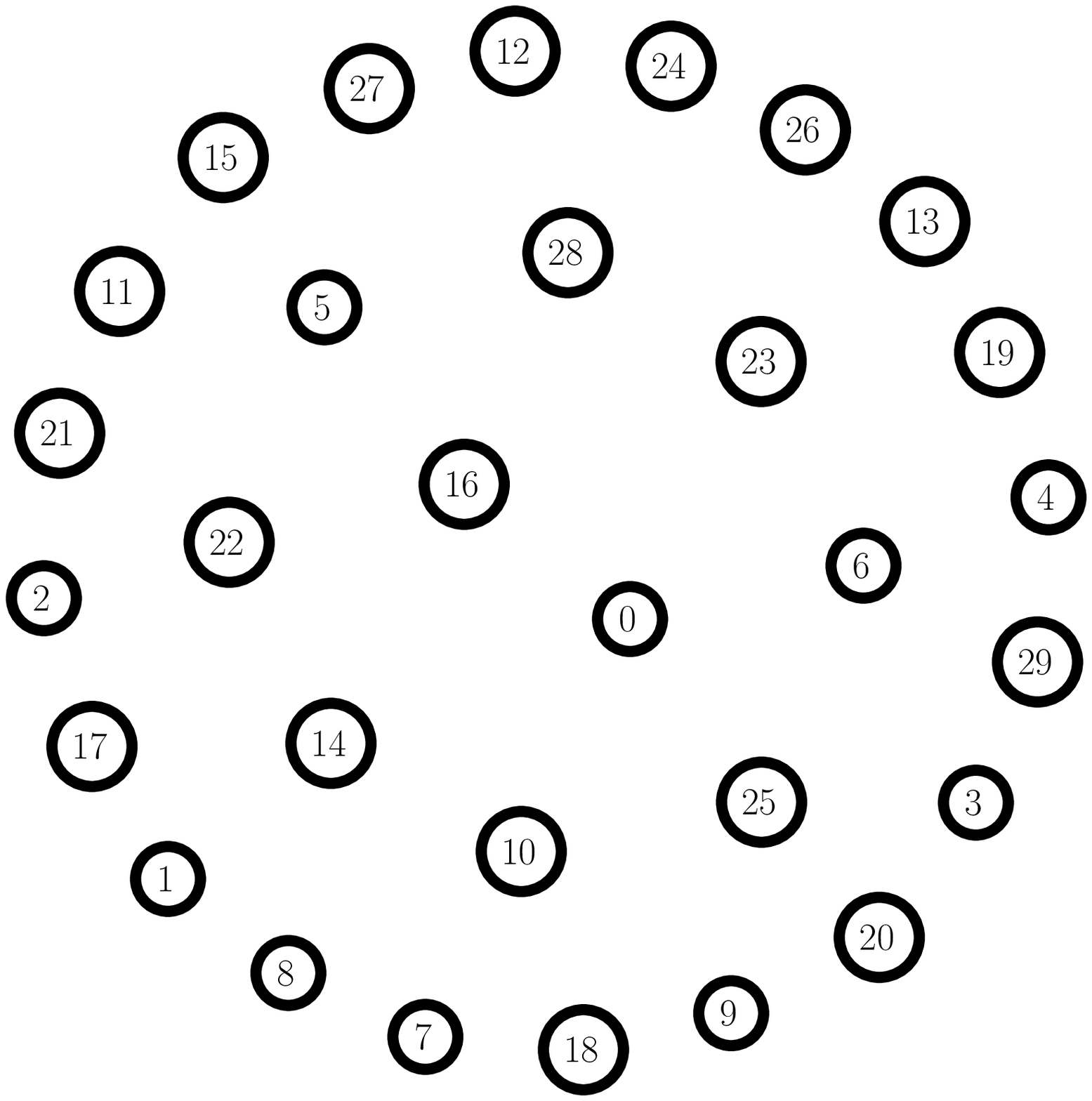}}}
      \cnodeput*(0,8){B}{ \scalebox{0.25}{\includegraphics[angle=0] 
          {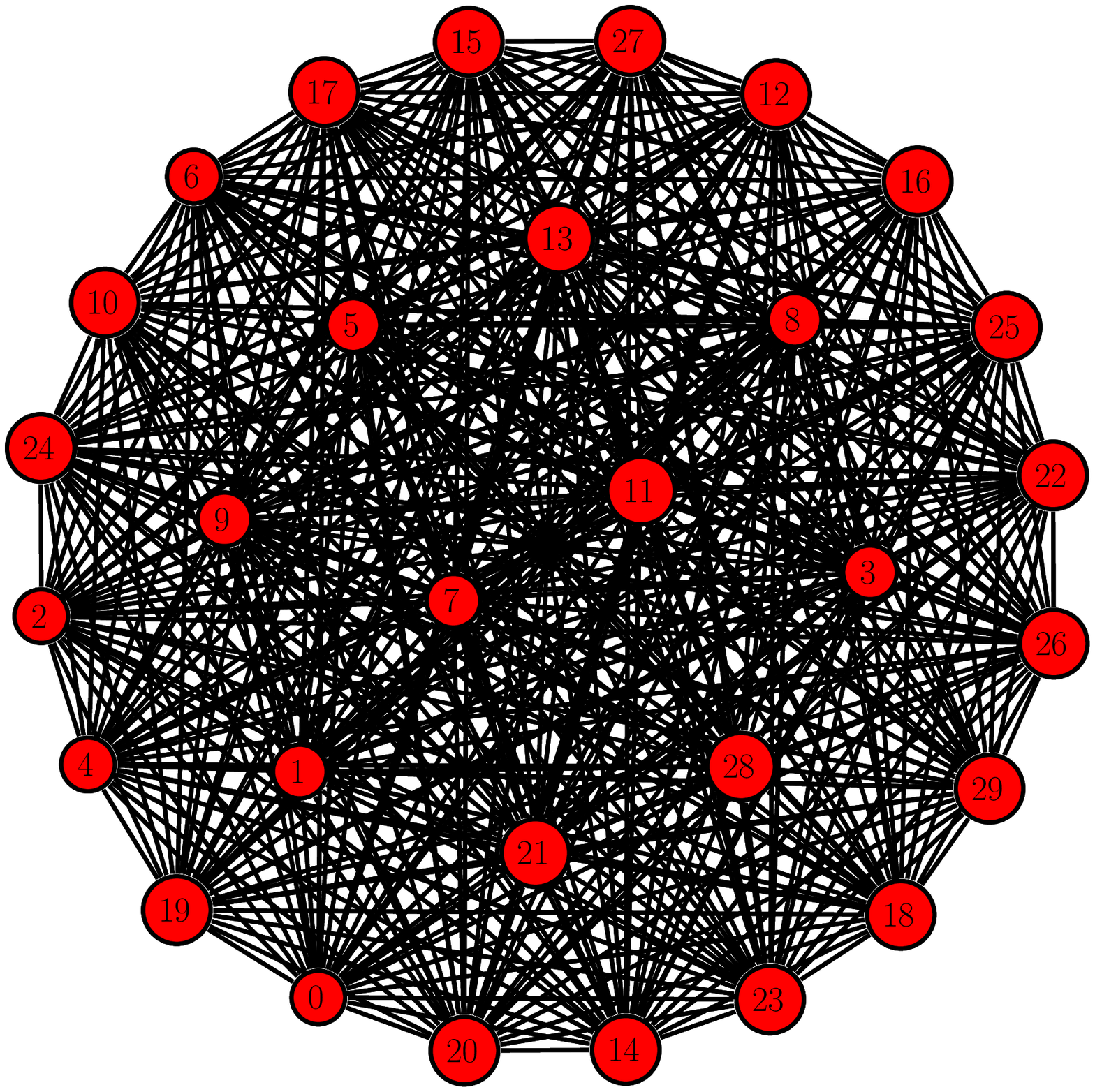}}}
      \cnodeput*(5,2){C}{ \scalebox{0.35}{\includegraphics[angle=0]
          {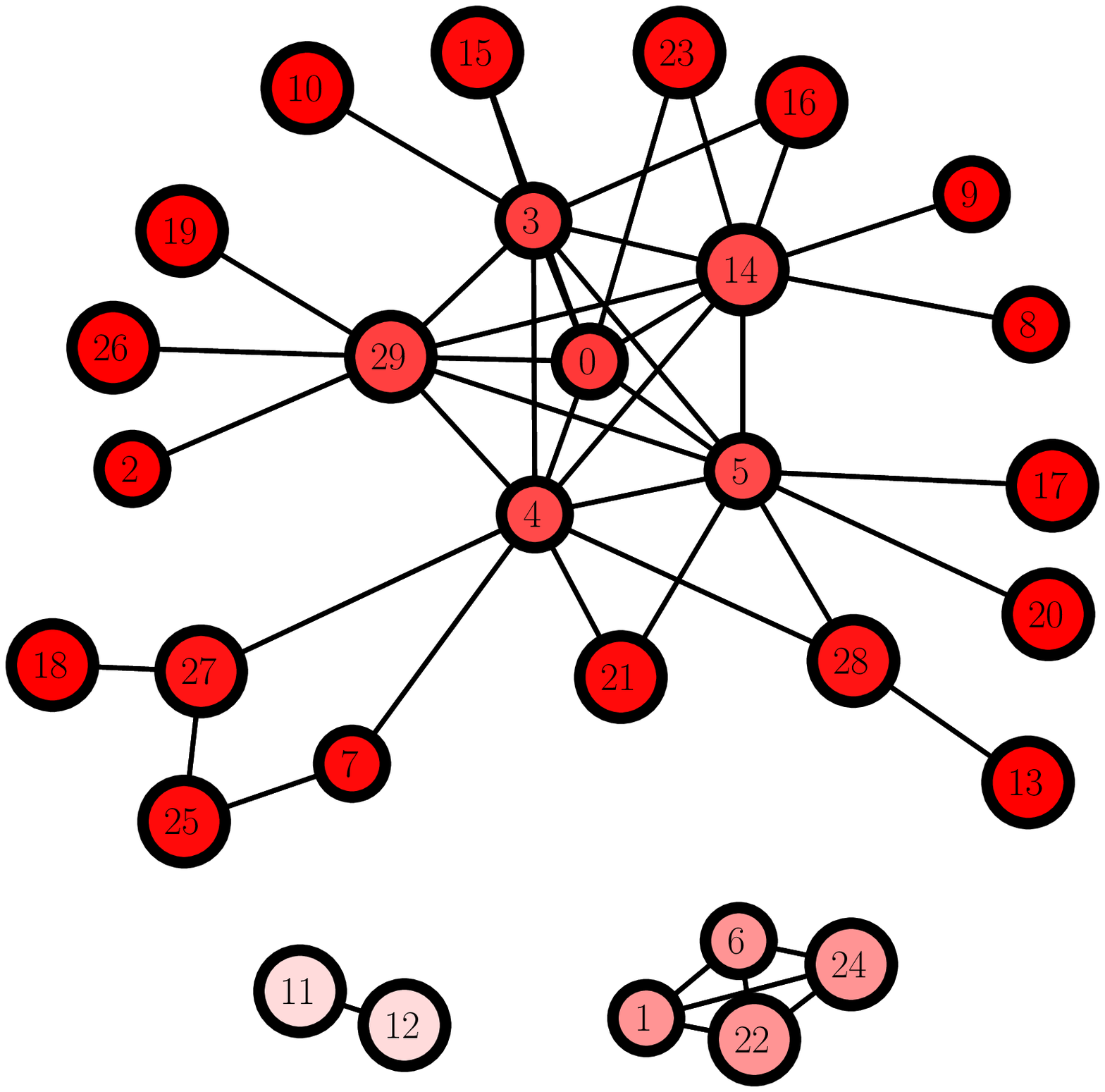}}}
      \cnodeput*(10,8){D}{ \scalebox{0.275}{\includegraphics[angle=0]
          {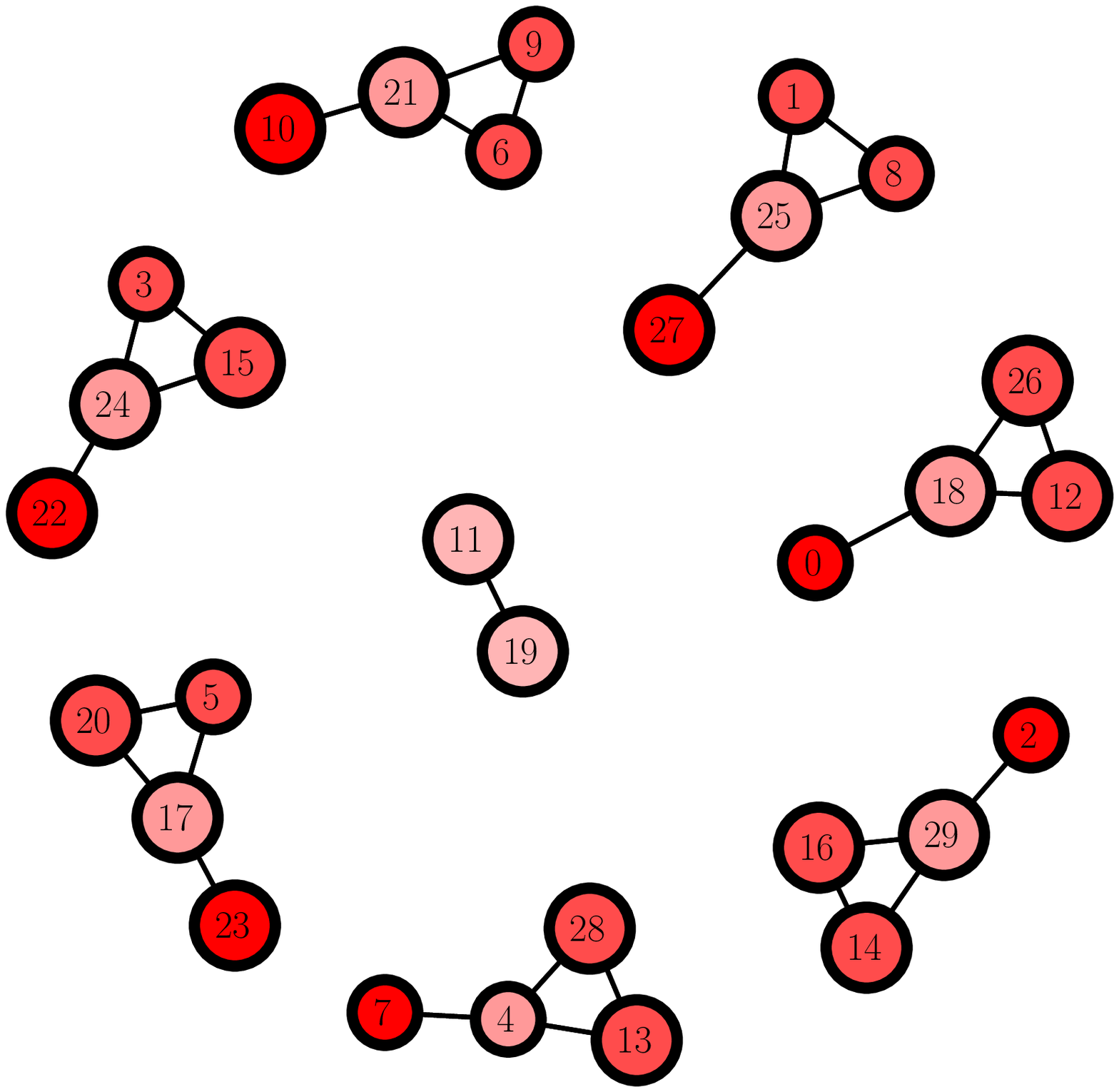}}}
      \cnodeput*(2,10){B2}{}
      \ncline[arrowsize=5pt 1.5]{->}{A}{B2} 
      \ncput*{$c=0.01$} 
      \cnodeput*(5,6){C2}{}
      \ncline[arrowsize=5pt 1.5]{->}{A}{C2}
      \ncput*{$c=0.2$} 
      \cnodeput*(8,10){D2}{}
      \ncline[arrowsize=5pt 1.5]{->}{A}{D2} 
      \ncput*{$c=0.5$}
    \end{pspicture}
  \end{center}
  \caption{Initial (empty) network (top) and stable equilibrium networks
    for costs $c=0.01$ (left), $c=0.2$ (bottom) and $c=0.5$ (right). The
    links indicate the mutual exchange of knowledge between agents (R\&D
    collaborations). For all values of cost the complete graph is most
    efficient, but only for a very small cost, $c=0.01$ (left), it is
    reached in the network evolution. For intermediate costs, (bottom),
    those agents with a high degree, that are maintaining many links,
    have smaller utility than those with a small degree.  The color
    saturation of the nodes indicate the utility of the agent compared to
    the maximum utility.  For high costs the stable equilibrium graph is
    identical to the initial empty graph.  }
  \label{fig:graph_plots}
\end{figure}

In Fig. (\ref{fig:graph_evolution_c_0.2}), the evolution of the network
measures mentioned above (the network density, the relative performance,
average degree and the global clustering coefficient) is shown for some
particular realizations with $n=30$ agents and different values of cost,
$c \in \{0.01,0.2,0.5\}$, $c \le 0.5$.  The values we measure are
relative quantities with respect to the complete graph which is in this
reange of cost also the efficient graph. One can see that for $c=0.2$ and
$c=0.5$, the stable equilibrium network is inefficient, sparse and highly
clustered. It is important to notice that those agents with high degree,
which bear the cost of many interactions, have smaller utility than those
with a smaller degree.  This is also indicated by the colors of the nodes
in Fig. (\ref{fig:graph_plots}).  The agents with small degree are
benefiting to a larger extent than the high degree agents. This comes
from the properties of the largest real eigenvalue of the adjacency
matrix. The eigenvalue of the network, which determines the positive
contribution to the individual growth rates, is the same for all the
agents in the same component, but the costs are depending on the degree.
Accordingly, the nodes with high degree have the same return as the nodes
with small degree from the network but they have to incur higher costs.

In Fig. (\ref{fig:graph_plots})\footnote{The graphs were plotted with a
  network layout algorithm introduced by
  \citet{geipel07:_self_organ_dynam_networ_layout}.} the stable
equilibrium networks for $30$ agents and three different values of the
cost are shown. For small costs, $c=0.1$, the complete and efficient
graph is reached. For intermediate costs, $c=0.2$, a sparse and highly
clustered graph with a highly heterogeneous degree distribution is
obtained. For high costs, $c=0.5$, the stable equilibrium network
consists of many small clusters. One can see that by decreasing the cost
the size of the connected components grows. This is consistent with what
has been observed in a recent study by
\citet{hanaki07:_dynam_r_d_collab_it_indus} on R\&D collaborations of
firms in the IT industry.

\begin{figure}[htpb]
  \centering
  \begin{minipage}{.45\linewidth}
    \psfrag{network density}[c][][3][0]{$s$}
    \psfrag{tau}[c][][3][0]{$\tau$}
    \centerline{\scalebox{0.4}{\includegraphics[angle=0]
        {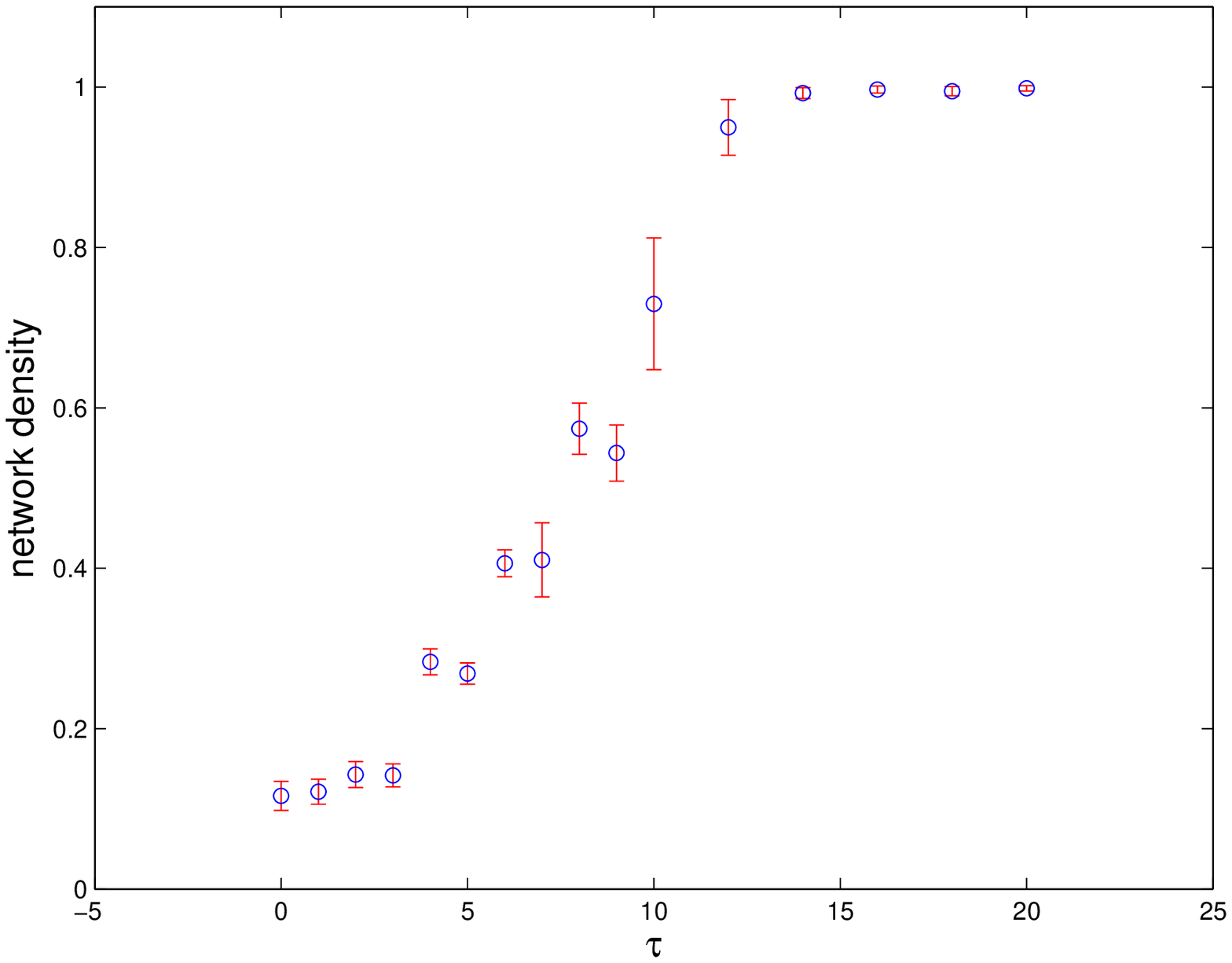}}}
    \bf{a}
  \end{minipage}
  \hfill
  \begin{minipage}{.45\linewidth}
    \psfrag{relative total growth}[c][][3][0]{$\pi$}
    \psfrag{tau}[c][][3][0]{$\tau$}
    \centerline{\scalebox{0.4}{\includegraphics[angle=0]
        {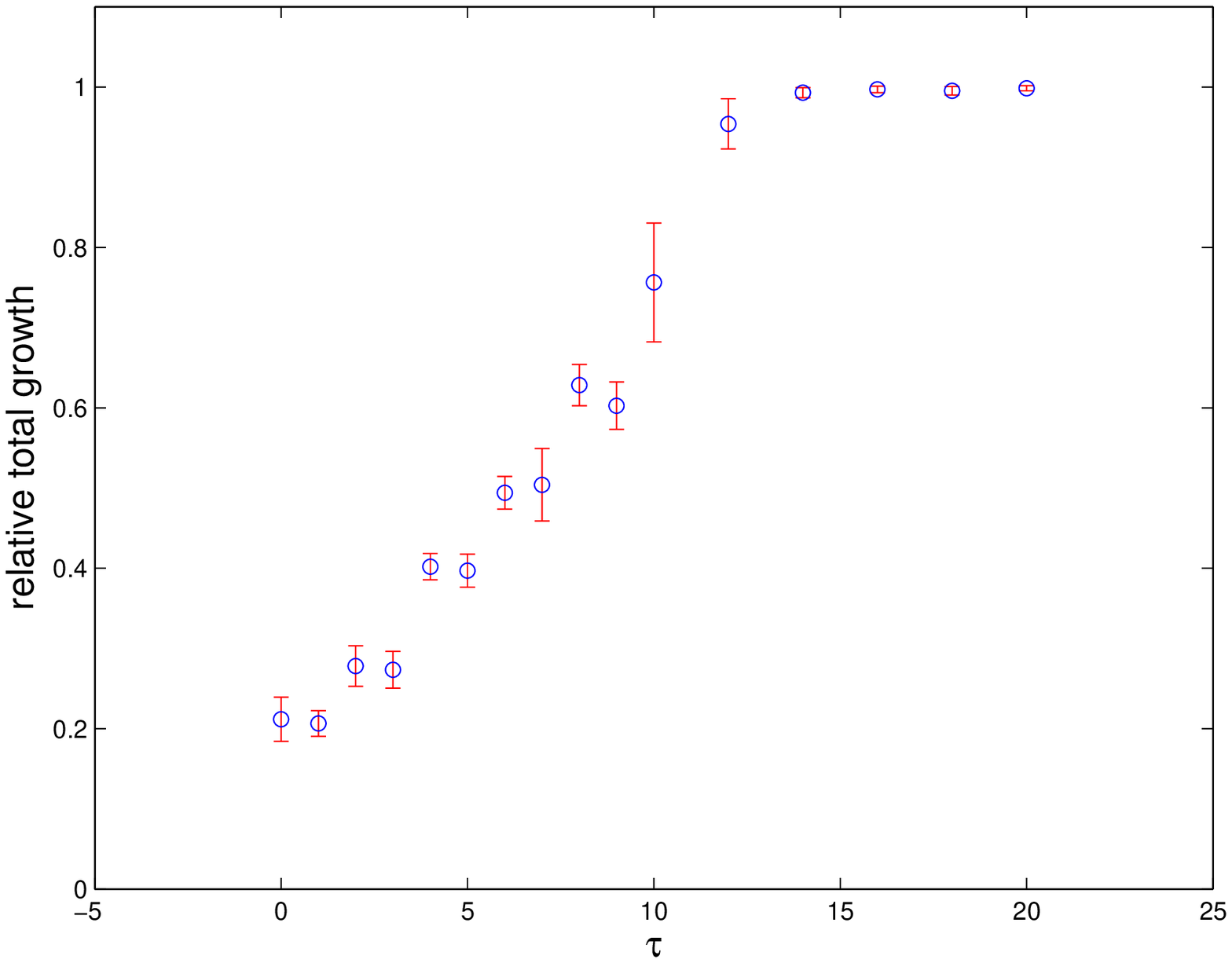}}}
    \bf{b}
  \end{minipage}
  \caption{Density (\textbf{a}) and relative performance
    (\textbf{b}) for the stable equilibrium networks with cost $c=0.2$,
    $n=30$ agents and $10$ realizations for every evaluation period $\tau
    \in [0,20]$.  By increasing the evaluation period $\tau$, the
    efficient graph, $K_n$, is reached.}
  \label{fig:expiration_times}
\end{figure}

In Fig. (\ref{fig:expiration_times}) network density $s(G)$ and relative
performance $\pi(G)$ are shown for increasing values of the evaluation period
$\tau$ ($10$ realizations for every value of $\tau$) for $n=30$ agents
and intermediate costs $c=0.2$. If the evaluation period is long enough,
the efficient graph, i.e. the complete graph, can be reached. Thus, if
agents are evaluating their interactions in the long-term, the
performance of the system can be increased up to the efficient state.

\section{Conclusion}
\label{sec:discussion}

In this paper, we consider the economic model of network evolution
introduced in \citep{koenig07:_effic_stabil_dynam_innov_networ} in the
context of innovation and R\&D collaborations. The model is characterized
by two time scales: there is a fast dynamics on the state variable of the
nodes, representing their knowledge, and a slow dynamics on the links of
the graph. Since the fast dynamics is linear and occurs on a static
graph, there is a number of well known results form the theory of
matrices that can be applied to the model and we have reviewed the most
important of them. For what concerns the evolution of the network we have
used some results from the theory of graph spectra to derive some
propositions on the efficiency and stability of the network. In
particular, we have provided a simple proof of the existence of
equilibria, like the star and the clique, that for $c<1/2$ are not
efficient but are stable. The existence of inefficient equilibria is of
interest to economists because it raises the issue of how to design
appropriate policies to help the system to reach the efficient
equilibria. Our simulations confirm the analytical results and show that
the interplay between dynamics on the nodes and topology of the network
leads in many cases to equilibrium networks which are not efficient and
are characterized, as observed in empirical studies of R\&D networks, by
sparseness, presence of clusters and heterogeneity of degree. In
particular, we observe subgraphs of finite size and highly heterogeneous
degree distribution among which there are only a few connections. These
properties have been observed in empirical studies of innovation networks
\cite{cowan04:_knowl_dynam_networ_indus,hanaki07:_dynam_r_d_collab_it_indus}
(for a more systematic comparison with the stylized facts on innovation
networks, see \citep{koenig07:_effic_stabil_dynam_innov_networ}).

As an new element, in this paper we also introduce a time $\tau$ after
which agents evaluate whether to keep or delete a link and we investigate
by means of computer simulation how the equilibrium reached by the
network is affected by increasing the time $\tau$. If agents evaluate
their interactions on a long-term, then they are able to reach an
efficient state, which, on the other hand, is not reachable, when
collaborations are evaluated in the short-term.  In other words, a
short-sighted rational behavior in the agents can give rise to
inefficient networks, as often happens in reality. Appropriate policy
measures could be designed to support economic agents in maintaining
interactions even when, in the short run, they may be unprofitable.  Our
model may serve as a first step towards both a theoretical explanation
for the empirical regularities in a R\&D networks and a very simple test
bed for policy design.

\section{Acknowledgments}
\label{sec:acknowledgements}

We are particularly grateful to Hans Haller for his astute comments and
perceptive suggestions. Morover we are in debted to Koen Frenken, Giorgio
Fagiolo, Matthias Feiler, Kerstin Press, Jan Lorentz and Nicolas Carayol
for clarifying discussions and helpful comments. Finally we would like to
thank the critical audiences at the 5th International EMAEE Conference on
Innovation in Manchester, 2007, and the audience at the European
Conference on Complex Systems in Dresden, 2007.

\bibliography{NHM-save}

\end{document}